\let\new=\newcommand
\new{\diff}{{\rm d}}
\new{\cE}{{\cal E}}
\new{\Jm}{{J^2_{max}}}
\new{\bp}{\mbox{\boldmath $p$}}
\new{\bx}{\mbox{\boldmath $x$}}
\new{\bv}{\mbox{\boldmath $v$}}

\input{epsf}

\documentstyle[11pt,aaspp4]{article}

\slugcomment{Submitted to \apj}

\lefthead{Mangalam, Nityananda \& Sridhar}
\righthead{Constrained violent relaxation to a spherical halo}

\begin{document} 

\title {Constrained violent
relaxation to a spherical halo} \author{ A. MANGALAM \\
mangalam@iucaa.ernet.in} \author{R. NITYANANDA \altaffilmark{1} \\
rajaram@rri.ernet.in} \and \author{S. SRIDHAR \\ sridhar@iucaa.ernet.in}
\affil{ Inter University Centre for Astronomy and Astrophysics \\ PostBag
4, Ganeshkhind, Pune 411 007, INDIA} \altaffiltext {1}{Permanent address:
Raman Research Institute, Sadashivanagar, Bangalore 560 080, INDIA}

\begin{abstract} 

Violent relaxation during the collapse of a galaxy halo is known to be
incomplete in realistic cases such as cosmological infall or mergers. We
adopt a physical picture of strong but short lived interactions between
potential fluctuations and particle orbits, using the broad framework
outlined by Tremaine (1987) for incorporating incompleteness of the
relaxation.  We are guided by results from plasma physics, viz. the
quasilinear theory of Landau damping, but allow for significant
differences in our case. Crucially, wave particle scattering does not
drive the system to an equilibrium distribution function of the
exponential type, even in regions of phase space allowed by the
constraints. The physical process is mixing without friction in ``action''
space, for which the simplest possible model is a {\em constant} phase
space density modulated by the constraints. Our distribution function does
not use the exponential functions of the energy prevalent in other work,
which we regard as inappropriate to collisionless systems. The dynamical
constraint of a finite short period of the relaxation process is argued to
lead to a $1/T_r$ factor in the distribution function, where $T_r$ is the
radial period.  The notion of strong potential fluctuations in a core is
built in as a cutoff in pericenter (which we find preferable to one angular 
momentum, the other
alternative we explored). The halo of the
self-consistent, parameter-free solutions show an $r^{-4}$ behavior in
density at large $r$, an $r^{1/4}$ surface brightness profile in the region
$0.1-8 ~r_e$, and a radially anisotropic velocity dispersion profile
outside an isotropic core. The energy distribution seen in simulations
$N(E)$ singles out the pericenter cutoff model as the most realistic among
the variants we have explored. The results are robust to modifications of
the period dependence keeping the same asymptotic behaviour, or to the use
of binding energy raised to the power of three halves in place of $1/T_r$.
     
\end{abstract} 
\section{A model of quasi-violent relaxation} 
\label{quasi} 

Elliptical galaxies are expected to have undergone violent relaxation
(Lynden-Bell 1967), which is a collisionless process whereby the energies
and angular momenta of stars and dark matter particles get redistributed
by strong potential fluctuations in such a way that the outcome depends
mainly on the macroscopic features of the initial conditions. This concept
is supported by N-Body experiments where for a range  of initial
conditions, the final state has more or less a universal profile.

One of the insights gained from cold collapse simulations is that a
compact region or a core develops, the system partially reexpands and then
in a few crossing times settles into a centrally peaked configuration with
most of the mass inside a radius within a fraction of the initial size (van
Albada 1982). The potential fluctuations are initially of large amplitude
but they damp out in a few crossing
times. Clumpy and cold initial conditions with small values of $2T/W
\simeq 0.1$, where $T$ and $W$ are total kinetic and potential energies,
are preferred over hot or homogeneous conditions in order to produce final
configurations that fit the de Vaucoleurs $\exp (-r^{1/4})$ surface
density profile better.

The general violent relaxation problem is difficult because the
fluctuations have large amplitudes and damp in a few crossing times. Thus
there is no small parameter and just a single time scale.  The physical
process belongs to the broad category of interactions between {\em waves}
(potential fluctuations) and {\em particles} (stars or dark matter),
wherein the amplitude of the wave is self-consistently determined by the
positions of the particles. We start with the limiting case of small
amplitudes, and long-lived waves, where we might expect the problem to
submit to perturbative methods. However, it is difficult to determine even
the linear modes of self-gravitating, collisionless systems; this is
because galaxies are spatially inhomogeneous objects.  Quasi-neutral
plasmas are not subject to this problem, so let us examine the
quasi-linear theory of Landau damping of electrostatic waves in
collisionless plasmas (c.f. \S~49 of Lifshitz \& Pitaevskii~1981). The
focus of the theory is on the slow evolution of the coarse-grained DF
(distribution function) of electrons, $f(\bp, t)$, where $\bp$ is
momentum. A key result is that $f(\bp, t)$ obeys a diffusion equation in
$\bp$--space:

\begin{equation} 
\frac{\partial f}{\partial t} =
\frac{\partial}{\partial p_{\alpha}} \left( D_{\alpha\beta}\frac{\partial 
f}{\partial
p_{\beta}}\right)\,, \label{diff} 
\end{equation} 
\noindent 
where the diffusion coefficient, $D_{\alpha\beta}$, is nonzero only in a
range of $\bp$ corresponding to electrons that are resonant with the
perturbations. This diffusion causes smoothing, and creates a plateau in
$f(\bp)$. Furthermore, $D_{\alpha\beta}$ is proportional to the energy in
the fluctuations; as this energy is absorbed by the resonant electrons,
the diffusion itself weakens, and $f(\bp)$ reaches a steady state. The
point we wish to make here is that this evidently non-Maxwellian steady
state has been reached by a self-regulating diffusive process.

There is of course a basic difference between slow damping of plasma
waves, and the damping of the oscillations of a violently relaxing galaxy.
In the former, the energy in the waves is absorbed by resonant electrons,
whereas the very concept of resonance is dubious in the latter case,
because the oscillations are short-lived. In a relaxing galaxy, the orbits
of particles are expected to be scattered by an oscillating core (Tremaine
1987, Spergel \& Hernquist 1992 (SH)). In realistic situations, the
deflections are individually large but small in number. Let us,
nevertheless, for orientation, initially consider the limit in which each
star, or dark matter particle, undergoes many small kicks. Hence particles
move under the combined actions of integrable forces, and small kicks that
lead to global stochasticity. In stochastic regions of phase space far
from islands, the diffusion of actions is well described by a
Fokker--Planck (FP) equation. A feature of the FP equation for Hamiltonian
kicks is the complete absence of dynamical friction, a result as old as
Landau~1937 (c. f. \S~5.4 of Lichtenberg \& Lieberman~1983). The FP
equation resembles equation~(\ref{diff}), where $\bp$ are the actions.
Once again, $D_{\alpha\beta}$ is proportional to the square of the
perturbation.  The absence of friction implies that a Maxwellian
distribution is not the final steady state. In fact, the process being
purely diffusive, it is evident that the DF would approach one that is
independent of the actions, given a finite volume of phase space and
enough time.

 We learn from the above examples that in the limit of long-lived, small
amplitude fluctuations, the relaxation of a galaxy is likely to be
primarily diffusive in phase space.  This relaxation process is very
different from collisional relaxation in neutral gases, which is a
reflection of the long-range nature of gravitational forces in contrast to
the short range of interatomic forces. The distinction is made clearer by
using the H-functions introduced by Tremaine, Henon, and
Lynden-Bell~(1986). They are actually functionals of the coarse-grained
DF, defined as 

\begin{equation}
 H[f]=-\int
C(f)\diff^3x\,\diff^3v\,,\label{hfunc} 
\end{equation} 
\noindent 
where $C(f)$ is a convex function. Any mixing process conserving fine
grained phase density, such as collisionless relaxation, will increase
$H[f]$. This result needs the assumption that the initial state had the
fine and coarse grained distribution equal to each other, which applies to
the cold collapse simulations with which we are concerned with. It does not
imply that $H$ increases monotonically (Dejonghe 1987, Sridhar 1987). On
the other hand, binary collisional processes, such as those relevant to
thermal relaxation in neutral gases, do not conserve phase density and
single out a unique H-function, namely the Boltzmann entropy given by
$C(f)=f\ln{f}$.

We now extrapolate to the case of the large amplitude, short--lived
fluctuations appropriate to violent relaxation. When the ``actions''
change by large amounts, the mixing process is no longer correctly
described by a diffusion equation. However, the relaxation is probably
well described by a process of mixing without friction, and an initial
DF will spread as far as it can in phase space; the extent and nature of
the spreading being determined by dynamical constraints discussed in the
following section. We write the DF~$=A(\cE)\,\times$ factor expressing
constraints, where $\cE$ is the single particle binding energy (bound
particles have positive $\cE$).  In the short duration before freeze-out,
$A$ spreads out nearly evenly and beyond $\cE=0$ (some particles escape).
But in constructing distribution functions of the galaxy we do not include
unbound particles. 
The parallel to
diffusion without friction suggests a final state for which $A(\cE)={\rm
constant}\,$.

Note that this hypothesis is the simplest choice consistent with the
physics of violent relaxation. A similar concept was proposed by Yan'kov
(1994) to explain turbulent transport in plasmas of tokomaks with a
uniform distribution of particles in a region specified by certain
constraints of the problem. 

\section{A prescription for $f(\cE,J^2)$} 
\label{guess} 
The relevant
dynamical constraints will clearly depend on the context;  for instance,
the merger of two galaxies is very likely to be different from the
formation of dark halos by cosmological infall over extended periods of
time. We now seek the dynamical constraints appropriate to situations such
as the cold collapses familiar from numerical simulations (c.f. van
Albada~1982) which might also carry over to the case of head-on mergers
between galaxies. For simplicity, we assume that the relaxed system is
spherical, and its DF is a function of the energy, $\cE$, and the angular
momentum, $J$. Tremaine~(1987)  has suggested the following two plausible
physical requirements (see also Merritt, Tremaine, \& Johnstone~1989
(MTJ), Stiavelli \& Bertin ~1987).

 \noindent {\bf R1.} The potential fluctuations last for only a limited time, 
$T_e$, which is of order
a few crossing times. Hence orbits with radial periods, $T_r$, exceeding $T_e$ 
will 
be underpopulated by a factor $\sim (T_e/T_r)$. In terms of a plausible physical 
requirement, this is viewed as a uniform filling of those
orbital phases which allowed the particle to visit the core in the time
$T_e$, and zero elsewhere. Finally, of course, all phases are equally
populated in the coarse grained function, which means a filling
proportional to $T_e/T_r$.

\noindent {\bf R2.} The fluctuations are largely confined to a central
region or a core of radius $r_c$. Hence only orbits whose pericenters are
smaller than $\sim r_c$ would have visited the region of large potential
fluctuations, and have had a chance to relax violently. The validity of
this will depend strongly on the initial conditions; the collapse of a
cold, nearly non-rotating initial configuration is more relevant to this
study.

With this motivation, we assume the following form for $f(\cE,J^2)$

\begin{equation}
f(\cE, J^2) = \left ( \begin{array}{c} \mbox{A function proportional} \\
\mbox{to $1/T_r $}\\ \mbox{for large $T_r$} \end{array} \right )  \times
\left ( \begin{array}{c} \mbox{Cutoff in pericenter} \\ \mbox{or} \\
\mbox{in $J^2$} \end{array} \right ) \times A 
\label{us}
\end{equation}
 
The first two factors in the distribution function are the dynamical
constraints given by {\bf R1} and {\bf R2} respectively, and the third
factor, $A$, is taken to be a constant as discussed earlier.  The first
factor $1/T_r$ ensures the continuity of the DF at $\cE=0$, i.e., $f(\cE,
J^2)=0$ for $\cE \leq 0$.  Jaffe (1987) remarked that one asymptotic
property of violently relaxed systems is that $N(\cE)$ should be finite
and non-zero at $\cE=0$ because the number of particles ejected from the
core should be smooth across $\cE=0$.  We show in \S \ref{states} that for small 
$\cE$, the restricted density of states due
to the second factor in (\ref{us})  is $\sim \cE^{-3/2}$, while $f \sim
1/T_r \sim \cE^{3/2}$ (for finite mass systems) and therefore $N(\cE) \sim
\cE^0$. We define 
\begin{equation} 
f(\cE, J^2) \equiv f_0(\cE, J^2) \cdot {\cal C}
\cdot A 
\end{equation} 
where $f_0$ and $\cal C$ represent the first and second
factors in (\ref{us}). In this work, we studied the following closely
related models based on (\ref{us}).

\begin{itemize}

\item
A model with $f_0(\cE)=\cE^{3/2}$  was first 
obtained and used as an input to generate the $1/T_r$ counterpart discussed 
below. Details of this model are presented in appendix \ref{e1.5}. The   
dependence of the distribution function on $\cE$ and $J$ is explicit, and hence 
the analytic work can be carried further. 

\item 
\label{trpc}
$f_0(\cE,J^2)=1/T_r$, is presented in \S \ref{1/tr}. Here, the dependence
of the radial period on  $\cE$ and $J$ has to be derived
iteratively from the potential. The
models above form a good starting point for  successive
numerical iterations which  we refer
to  as $r_c$ and $J_m$ models.  A variant of these models with
$f_0(\cE,J^2)=1/(T_r+T_0)$,
with an extra parameter $T_0$ introduced to assess the sensitivity of the
results to the functional form of the period cutoff.

\end{itemize}

All the above models have the asymptotic properties $\rho \sim r^{-4}$,
$N(\cE) \sim \cE^0$ (described above), and surface densities approximating
the $r^{1/4}$ law. The form of energy distributions, $N(\cE)$, at
distances larger than the core radius $r_c$, or the scale radius $r_j$,
are derived for the  pericenter model in \S \ref{states}. In \S \ref{netr}, we 
describe the form of $N(\cE)$ for the
$T_r$ models. This is helpful in picking out the pericenter cutoff models
as a better fit to results of simulations.  In \S \ref{ftr} we describe
the boundaries in the $\cE-J^2$ plane required in doing the integrals in
the allowed velocity space.  
The properties of the solutions
and comparison to simulations are described in \S \ref{prop}, and we
present a discussion in \S \ref{disc}, and conclusions in \S
\ref{conclude}. The reader interested in first examining the results is
urged to go to \S \ref{prop}--\ref{conclude} and then return to \S
\ref{ftr}--\ref{1/tr} for some details of the construction.

\section{Pericenter and angular momentum constraints in $\cE-J^2$ plane}
\label{ftr}
For a spherical model, with a distribution function, $f(\cE,J^2)= A 
f_0(\cE,J^2)~ {\cal C}$, including a sharp pericenter cutoff at $r_c$ or in 
angular momentum, $J_m$, Poisson's equation can be written as
\begin{equation}
 {1 \over r^2} {\diff \over \diff r} (r^2  {\diff \Phi \over \diff r}) 
= {A \over r^2}\int \diff \cE ~\diff J^2 (2 (-\cE-\Phi) -J^2/r^2)^{-1/2}~ 
f_0(\cE,J^2)
~{\cal C}(\eta),
\label{poisson}
\end{equation}
 where $A$ is a constant that is determined by normalization. The  cutoff 
function, ${\cal C}$, is given by
\begin{equation} 
{\cal C}(\eta)= \left\{ \begin{array}{ll} 1 &  \eta \leq 1 \\ 0  & \eta > 1 
\end{array} \right.
\label{cutoff}
\end{equation}
where $\eta$ equals  either $r_p/r_c$ where $r_p(\cE, J^2)$ is the pericenter, 
and $r_c$ is the cutoff radius, or  $J/J_m$, where $J_m$ is the maximum angular 
momentum. We explore both these constraints here  and construct a method to 
solve the self-consistent models with $f_0(\cE, J^2)= 1/T_r$; eqn 
(\ref{poisson}), coupled with
\begin{equation}
T_r(\cE, J^2) = 2~ \int_{r_p}^{r_a} \diff r~ (2 (-\cE-\Phi)
-J^2/r^2)^{-1/2}\,,
\end{equation}
where $r_p$ and $r_a$ are the turning points. We have solved the coupled 
equations by a semi-analytical method whose details are presented in \S 
\ref{1/tr}. 

Now we introduce a variable, $r_0(J^2)$, which is the radius of a circular
orbit for a given angular momentum, $J^2=r_0^3 \Phi'_0$. This is useful in
determining the region of integration in the $\cE-J^2$ plane. Note that
$J^2= M(r_0) r_0$ from Poisson's equation, where $M(r)$ is the mass inside
$r$ and hence it is a monotonically increasing function of $r_0$. 
Fig. \ref{ro} shows a plot of the absolute value of
the effective potential, $-\cE_f(r,r_0) \equiv \Phi +J^2 /(2 r^2)= \Phi
+\Phi'_0 {r_0}^3/(2 r^2)$ for a given $J^2$ and a typical $ \Phi$. We now
consider the region of integration allowed by ${\cal C} (\eta)$.

\subsection{Case of $\eta =r_p/r_c$}
\label{ftr1}

The region of integration in the $\cE -J^2$ plane is bounded by following curves 
which are shown in Fig. \ref{ej2}.

\noindent {\bf K1.} The upper bound of the region of interest is given by the 
minimum of the effective potential ($-\cE_f$) of a circular orbit for a given 
$\Phi(r)$ and $r_0$ ie, $\cE < \cE_f(r_0,r_0)= -\Phi_0 -\Phi'_0 r_0/2$ for $r_p 
\leq r_0 \leq r_c$. The orbits with $r_p \leq r_c \leq r_0$,  obey the bound, 
$\cE < \cE_f(r_c,r_0)=-\Phi_c -\Phi'_0 {r_0}^3/(2 r_c^2)$ as indicated in Fig 
\ref{ro}. 

\noindent {\bf K2.} At a given radius $r$ for the system, the effective 
potential is bounded by $\cE \leq   B(r,J^2) \equiv \cE_f(r,r_0)=-\Phi-J^2/(2 
r^2)$ or equivalently $v_r^2(r) \geq 0$. 

\placefigure{ro}
For $r<r_c$, the pericenter constraint is satisfied and the operative bound is 
just
$\cE < \cE_f(r,r_0)$. Hence, the bounding line $\cE = \cE_f(r_c,r_0)$ given by 
constraint  K2 lies inside and is tangent to the curve given by constraint  K1; 
the point of contact represents a circular orbit for a given $r_0$. 

Now consider the case $r> r_c$. The point of intersection, ($\cE_\ast, 
J^2_\ast$), between the bounding line of bound K2,  $\cE = \cE_f(r,r_0)$ and 
$\cE =\cE_f(r_c,r_0)$ given by bound  K1 represents an orbit whose turning 
points are $r$ and $r_c$ (see Fig \ref{ej2}). 
\placefigure{ej2}
The point of intersection works out to be
\begin{equation}
\cE_\ast=(\Phi_c r_c^2 -\Phi r^2)/(r^2-r_c^2)
\label{e*}
\end{equation}
\begin{equation}
J^2_\ast= 2 (\Phi  -\Phi_c ) r^2 r^2_c/(r^2-r_c^2)
\end{equation}

Effectively, for $r>r_c$, the bound given by constraint  K2,  $\cE < 
\cE_f(r,r_0)$, applies upto $J_c^2=  r_c^3 \Phi'_c$, beyond which  the bound, 
$\cE <-\Phi_c -J^2/(2 r_c^2)$ given by K1 is operative,  and forms a wedge 
shaped region. For $r <r_c$, the region of integration is given by the bound 
$\cE < \cE_f(r,r_0)$ and is a triangular shaped region. The regions of 
integration are shown in Fig. \ref{wedge} and are summarized by

\begin{equation}
{\cal A}_1 \equiv \cE < -\Phi-J^2/(2 r^2),  ~~~~~ r \leq r_c
\label{a1}
\end{equation}
\begin{equation}
{\cal A}_2 \equiv  \left\{ \begin{array}{lll} \cE < -\Phi-J^2/(2 r^2), & r>r_c  
&\& ~J^2 < J^2_\ast \\
\cE <-\Phi_c -J^2/(2 r_c^2) & r>r_c  &\& ~J^2 >J^2_\ast \end{array} \right.
\label{a2}
\end{equation}
\placefigure{wedge}
Now that the regions of integration have been determined, we can write the 
integral on the RHS of (\ref{poisson}), without the normalization constant, $A$ 
as
\begin{equation}
{\cal I}(r;f_0) = \left \{ \begin{array}{ll}   {\cal I}_0(r;f_0)  & r\leq r_c \\
 {\cal I}_0(r;f_0) - {\cal I}_-(r;f_0) +{\cal I}_+(r;f_0)
& r> r_c \\ \end{array}\right.
\label{i}
\end{equation}
where $J^2_e(r)=-2 \Phi r^2$ is the intercept on the $J^2$ axis, and 
\begin{equation}
 {\cal I}_0(r;f_0) = {1 \over r^2} \int_0^{J^2_e} \diff J^2 \int_0^{B(r,J^2)} 
\diff \cE~ ( 2(-\cE-\Phi) -J^2/r^2)^{-1/2} f_0(\cE,J^2)
\end{equation}
\begin{equation}
{\cal I}_-(r;f_0) = {1 \over r^2} \int_{J_\ast^2}^{J_e^2} \diff J^2 
\int_0^{B(r,J^2)} \diff \cE~ ( 2(-\cE-\Phi) -J^2/r^2)^{-1/2} f_0(\cE,J^2)
\end{equation}
\begin{equation}
{\cal I}_+(r;f_0) = {1 \over r^2} \int^{\cal E_\ast}_0 \diff \cE 
\int_{J_\ast^2}^{-(\cE+\Phi_c) 2 r_c^2} \diff J^2 ~( 2(-\cE-\Phi) 
-J^2/r^2)^{-1/2} f_0(\cE,J^2)
\end{equation}

\subsection{Case of $\eta=J/J_m$}
\label{ftr2}
 
The constraint K2 applies here and the maximum allowed angular momentum is 
$J_m$. There are no circular orbits allowed outside the radius $t$ given by 
$J_m^2 \equiv -2 \Phi_j r_j^2 = t^3 \Phi'(t)$. Also, for $r<r_j$ only constraint 
K2 applies and for $r>r_j$, the orbits with $J^2>J^2_m$ are excluded. Hence the 
region of integration, $\cal B$ is given by
\begin{equation}
\begin{array}{lllll}
{\cal B}_1 &\equiv& \cE < -\Phi-J^2/(2 r^2)  &&r<r_j  \\
{\cal B}_2 &\equiv& \cE  < -\Phi-J^2/(2 r^2) & \& J^2 < J^2_m & r>r_j
\end{array}
\end{equation}
For $\eta=J/J_m$, the integral on the RHS of (\ref{poisson}), without the 
normalization constant $A$ will reduce to
\begin{equation}
{\cal K}(r;f_0) =  \left \{ \begin{array}{ll}   {\cal I}_0(r;f_0)  & r\leq r_j\\
 {\cal I}_0(r;f_0) - {\cal K}_-(r;f_0)& r> r_j \\ \end{array}\right.
\end{equation}
where 
\begin{equation}
{\cal K}_-(r;f_0) =  {1 \over r^2} \int_{J_m^2}^{J_e^2} \diff J^2 
\int^{B(r,J^2)}_0 \diff \cE~ 2(-\cE-\Phi) -J^2/r^2)^{-1/2} f_0(\cE,J^2). 
\label{k}
\end{equation}

\section{Restricted density of states for a pericenter cutoff}
\label{states} 
Here we calculate the density of states for a model that has a sharp pericenter 
cutoff. The restricted density of states is given by 
\begin{eqnarray}
g(\cE)&=& \int \diff^3 r ~{\cal C} (r_p/r_c) {\diff^3 v \over \diff \cE}=8 
\pi^2\int_{\cal A} \diff r\int_0^\Jm \diff J^2/\sqrt{2(-\cE-\Phi)-J^2/r^2} 
\nonumber\\
&=& 16 \pi^2 \sqrt{2}\int_{\cal A} r \diff r ~\left (\sqrt{(-\cE-\Phi) 
r^2}-\sqrt{(-\cE-\Phi) r^2-\Jm/2} \right )
\end{eqnarray}
 where $\Jm$ is determined by constraints. Hereafter we work with units where 
$GM=1$. There is no contribution to $g(\cE)$ from $\cE >-\Phi$ and therefore  
for a given energy, $\cE$, only the region $r < r_\cE$ is accessible where 
$\Phi(r_\cE) = -\cE$. Consult Fig. \ref{wedge} for the allowed range of 
integration, ${\cal A}$. For $r<r_c$, the pericenter constraint (K1) is 
satisfied and therefore, $\Jm= J^2_e = 2 (-\cE-\Phi) r^2 $ as given by K2. Now, 
take the case when $\cE > \cE_c$ and $r> r_c$ shown in Fig. \ref{ec} where 
$\cE_c \equiv -\Phi_c -\Phi'_c r_c/2$, the energy of the circular orbit of 
radius $r_c$. Earlier, the point of intersection of $-\Phi-J^2/(2 r^2)$ and 
$-\Phi_c -J^2/(2 r_c^2)$ was defined to be $\cE_\ast, J^2_\ast$. Here, 
$\cE>\cE_c>\cE_\ast$, and hence the pericenters lie inside $r_c$ and $\Jm=J_e$ 
as given by constraint K2. Now consider the case $\cE<\cE_c$ and $r>r_c$ as 
illustrated in Fig. \ref{x1x2} for which $J^2_c=-2(\cE-\Phi_c) r_c^2 < 
J^2_\ast$. It is  clear that $\Jm= {\rm Min}(J_c^2, J_e^2)$ and if $J_c^2 < 
J^2_e$, the particle has a $\Jm=J_c^2$ because any higher angular momentum will 
include orbits whose pericenters lie outside $r_c$. This happens only for $r_c < 
r< r_I$ where $r_I(\cE,r_c)$ is the apocenter of an orbit for a given energy 
$-\cE$ and pericenter $r_c$. It is given by  
\begin{eqnarray}
\cE= { \Phi_I r_I^2 -\Phi_c r_c^2\over r_I^2 -r_c^2},
\label{rs}
\end{eqnarray}
which has two roots;  we seek the one for which $r_I>r_c$. 
\placefigure{ec}
\placefigure{x1x2}
To summarize:
\begin{equation}
\Jm(\cE,r)={\rm Min}(J_c^2, J_e^2)= 2\left \{ \begin{array}{lll} (-\cE-\Phi) 
r^2; & -\Phi > \cE> \cE_c, & r<r_\cE\\
               (-\cE-\Phi_c)r_c^2; &  \cE_c >\cE_\ast>\cE, & r_c<r<r_I<r_\cE\\
                (-\cE-\Phi) r^2; & \cE_c >\cE>\cE_\ast, & r_c<r_I<r<r_\cE 
\end{array}\right.
\end{equation}
  We may then write
\begin{equation}
g(\cE) = 16 \pi^2  \sqrt{2} \left ( -g_c(\cE)+\int_0^{r_\cE} r^2  
~\sqrt{-\cE-\Phi}~\diff r  \right )
\label{dstates}
\end{equation}
where 
\begin{equation}
g_c(\cE)=\left \{ \begin{array}{lll} \int_{r_c}^{r_I}\diff r~ r^2~\sqrt{
\cE_\ast-\cE} (1-{r_c^2 \over r^2})^{1/2} & \cE_c >\cE_\ast>\cE & 
r_c<r<r_I<r_\cE\\ 
0 & {\rm otherwise} &\end{array} \right.
\end{equation}
We  can now use the equation above to calculate the density of states and the 
energy distribution for a Keplerian potential, the asymptotics of which are 
applicable more generally for finite mass systems. After some algebra, the 
restricted Keplerian density of states (in units where $G M=1$) is given by
\begin{equation}
g_k(\cE) = \pi^3 \sqrt{2}\left \{ \begin{array}{ll} \cE^{-5/2} & \cE \geq 
\cE_c\\ 
 -4~\cE^{-3/2}~\Phi_c r_c^2 (1 -\cE \Phi_c r_c^2) & \cE < \cE_c \end{array} 
\right.
\end{equation}
where it is continuous at $\cE_c= 1 / (2 r_c)$. The energy distribution for 
$f_0(\cE) = A' \cE^{3/2}$, will then go as
\begin{equation}
N_k(\cE) =  A' \pi^3 \sqrt{2} \left  \{ \begin{array}{ll} \cE^{-1} & \cE \geq 
\cE_c\\ 
 4 \Phi_c r_c^2 (1 -\cE \Phi_c r_c^2)  & \cE < \cE_c \end{array} \right.
\label{nepc}
\end{equation}

Note that for any finite mass potential, the above equations are valid for
large radius ($r \gg r_c$) and only approximate in the inner region where
the realistic potential deviates from $1/r$. Clearly, as $r_c \rightarrow
\infty$, $\cE_c \rightarrow 0$, and one recovers the Keplerian form of
$g(\cE) \propto \cE^{-5/2}$ and if $r_c$ is zero then $g(\cE)$ vanishes as
expected. Near small $\cE$, the velocities become more radial and the
unconstrained Keplerian density of states, $\cE^{-5/2}$, is reduced by a
factor $v_t^2 \propto \cE$.  In our model the assumption of a sharp
pericenter cutoff at $r_c$ leads to $g(\cE) \propto \cE^{-3/2}$, near
$\cE=0$ and the choice of $f(\cE) = 1/T_r \sim \cE^{3/2}$, at small $\cE$,
based on the dynamical arguments made earlier, is consistent with the
required property of the ``break of $N(\cE=0$)'', or the finite and
non-zero value of $N(\cE=0)$. Jaffe(1987) made the interesting point that
the demand of a break in $N(\cE)= f(\cE) g(\cE)$ at $\cE=0$ for a
(unrestricted) Keplerian density of states, leads to $f(\cE) \sim
\cE^{5/2}$ near $\cE=0$ and as a result, the density behaves as $\Phi^4
\propto r^{-4}$.  The self-consistent $r_c$ models and the $J_m$ models
(see the following section) presented here are infinite-radius and
finite-mass models with an $r^{-4}$ density profile. We have checked that
the finite and non-zero $N(\cE=0)$ property also holds for the angular
momentum restricted density of states.

\section{$f_0=1/T_r$ model}
\label{1/tr}
The numerical solution of Poisson's equation for $f_0= 1/T_r$ follows closely 
the analytics in \S \ref{ftr}. The 
areas of integration in $(\cE, J^2)$ space are determined by  (\ref{i},\ref{k}). 
Starting with the initial guess of the potential given in appendix \ref{e1.5} or 
corresponding one for angular momentum cutoff, the radial period $T_r(\cE,r_0)$ 
is calculated by a root solving and integration routine within the bound given 
by $\cE <\cE_f(r_0,r_0)$ for $r_0 <r_c$ and  $\cE <\cE_f(r_c,r_0)$ for $r_0 > 
r_c$ in case of the $r_c$ model. Similarly, the bound  $\cE <\cE_f(r_0,r_0); r_0 
< r_{\rm max}(\cE)$ was used in the case of the $J_m$ model where $r_{\rm 
max}(\cE)$ is radius of largest allowed circular orbit. A lookup (interpolation) 
table for $T_r(\cE, r_0)$ was prepared. For a given radius, $r$, the regions of 
integration  are determined by $\cal A$ or $\cal B$ and then the distribution 
function $1/T_r$ is integrated to calculate the density profile using

\begin{eqnarray}
 \label{numeric} 
{1 \over r^2}{\diff \over
 \diff r} (r^2 {\diff \Phi \over \diff r}) &=& K \left \{ \begin{array}{ll}
{\cal I}(r; T_r^{-1}) & r_c ~{\rm model} \\
{\cal K}(r; T_r^{-1}) & J_m ~{\rm model} \end{array} \right.\\
[1em] 
T_r(\cE, J^2) &=& 2~ \int_{r_p}^{r_a} \diff r~ (2 (-\cE-\Phi)
-J^2/r^2)^{-1/2}\,, 
 \end{eqnarray} 
\noindent
where $r_p$ and $r_a$ are the turning points and the value of $K$ is chosen so 
that the total mass is 1 and $G=1$. The next iterate of the potential is then 
trivially obtained from
\begin{equation}
\Phi(\xi)= -\int_\xi^\infty \xi^{-2} {\cal M}(\xi) \diff \xi
\end{equation}
where ${\cal M}(\xi)$ is the mass fraction inside $\xi$ that is calculated from 
the density, the RHS of (\ref{numeric}). Here, $\Phi(\infty)$ is taken to be 
zero to be consistent with the initial guess of $\cE^{-3/2}$ for the radial 
period. 

The numerical code was used to verify the $\cE^{3/2}$ solution (the
initial guess) and vice versa. A high precision routine was used to
calculate the radial period for an arbitrary potential and this was tested
against the well-known forms for isochrone and Kepler potentials and found
to be precise to within a millionth of the correct value. The verification
of the analytics established the robustness of the numerics used. The
numerical scheme converges rapidly to a solution in a few iterations. This
indicates that the properties of these models are close to those of the
$\cE^{3/2}$ models. Fig. \ref{tre1.5} illustrates that the $r_c$ model is
nearly isochronic.  For many purposes, one could have been satisfied with
the $\cE^{3/2}$ models, which are much easier to implement than the
$1/T_r$ models, for exploratory work. But this is strictly with the
hindsight provided by our constructing the models in the first place.

\placefigure{tre1.5}

\subsection{The energy distribution}
 \label{netr}

 Let us now look at the energy distribution,
$N(\cE)$. For a spherical system with the distribution, $f(\cE, J^2)$, one
can write the differential energy distribution (see Binney
\& Tremaine, eqn 4P-11) as 
\begin{equation}
 N(\cE) = 4\pi^2\int f(\cE, J^2) T_r(\cE,
J^2)\,\diff J^2\,. 
\label{ne}
\end{equation}
 In the case of $f= A ~{\cal
C}(\eta)/T_r$, we obtain \begin{equation} N(\cE) = 4\pi^2 A~\int {\cal C}(\eta) \diff
J^2 = 4 \pi^2 A~J^2_\eta(\cE) \label{jeta} \end{equation} where $J^2_\eta(\cE)$ is
the constraint boundary in the $\cE-J^2$ plane. In the case of the $r_c$
model, the shape of the constraint in this plane (c.f.  K1, \S \ref{ftr1})
determines the energy distribution which is given by 
\begin{equation} N_c(\cE)= 4 \pi^2
A\left \{ \begin{array}{ll} t^3 \Phi'(t); ~~{\rm where} -\Phi(t)-t
\Phi'(t)/2= \cE ~~~& \cE > \cE_c\\ -2 (\cE+\Phi_c) r_c^2 & \cE < \cE_c
\end{array} \right. 
\label{nerc}
 \end{equation}
 where $\cE_c$ is the energy of a
circular orbit at $r_c$ as defined in \S \ref{states}, and similarly for
$J_m$ model (c.f \S \ref{ftr2}), we
obtain
 \begin{equation}
 N_j(\cE)= 4 \pi^2 A\left \{ \begin{array}{ll} t^3 \Phi'(t); ~~{\rm where}
-\Phi(t)-t \Phi'(t)/2= \cE & \cE > \cE_m\\ J^2_m & \cE < \cE_m \end{array}
\right. 
\label{nejm}
 \end{equation} 
where the energy $\cE_m$ of the largest allowed circular orbit is given by
\begin{equation}
\cE_m= -\Phi(t) - t^2 \Phi'(t)/2; ~~t^3 \Phi'(t)= J^2_m.
\label{em}
\end{equation}
The Keplerian limit of eqn (\ref{nerc}) and eqn
(\ref{nejm}) agrees exactly with the Keplerian limits of $\cE^{3/2}$
counterparts, when one
takes into account the fact that for a Kepler potential, $T_r= \pi GM
\cE^{-3/2}/\sqrt{2}$. Therefore $A/A'= \pi /\sqrt{2}$ in units where $G M
=1$. It is clear that, near small $\cE$, the form of $N(\cE)$ is linear
for the $r_c$ model and the slope is $-8 \pi^2 A r_c^2$ (positive in
$N(E$)) while the $N(\cE)$ is flat for the $J_m$ model.  
From the dynamical
arguments in \S \ref{guess} we expect  orbits with small $T_r$ to be
more populated than the ones with larger radial periods. However, we
have no reason to adhere to the strict $1/T_r$ form for small $T_r$. It is
worth checking how sensitive our results are to a modification in which
$f_0$ is flatter at small $T_r$ while retaining the $1/T_r$
asymptotic behavior. We thus try \begin{equation} f_0(E,J^2) = {1 \over T_r +T_0} \end{equation}
where $T_0$ is a parameter which is chosen to be of the order of the
radial period of the harmonic oscillator, $T_h$, near the bottom of the
well and given by \begin{equation} T_h= {\pi \over \sqrt{ \Phi''(0)}} \end{equation} Again, the
maximum deviation from the initial guess upto the second iterate in the
density was found to be less than $1 \%$ for $T_0= 0.5, 1, 2 ~T_h$. This
indicates that the distribution function is probably stable and
insensitive to small changes in $f_0$ but sensitive to the choice of the
constraint, ${\cal C} (\eta)$. This and the similarity of the $\cE^{3/2}$
models to the $1/T_r$ counterparts can be explained by the fact that $f_0$
is dominated by $\cE^{3/2}$ (see the contour plot of $T_r \cE^{3/2}$, Fig
\ref{tre1.5}).
  
\section{Properties of the solutions and comparison with simulations}
\label{prop}

 We discuss the analytic results in the preceding sections and
the properties of the $r_c$ and $J_m$ solutions and compare it with
relevant simulations and the $r^{1/4}$ law. As mentioned in \S \ref{quasi}
the simulations of direct relevance are cold collapse simulations, in
particular the C runs of van Albada (1982) who has presented the density,
$N(\cE)$, velocity dispersion and anisotropy profiles of the final
configurations. In our figures \ref{ro}, \ref{ej2}, \ref{wedge}, \ref{ec} and
\ref{x1x2}  a Henon isochrone potential
was used  to illustrate the allowed region in phase space.

\noindent {\it Scaled quantities and their physical units}

The models have two free parameters, the total mass $M$, and
either the cutoff radius $r_c$ or a maximum value of angular momentum, $J_m$.
Below, we compare the $r_c$ models with the $J_m$ models. The scale radius, 
$r_s$ is the scale radius of the $J_m$ model and equals the core radius, $r_c$, 
in the $r_c$ model. If the total mass and scale radius are set to unity, the 
model has no free parameters.

The potential in the $r_c$ solution is scaled according to 
\begin{equation} 
\Phi= {GM
\over a ~r_c }~\theta = 0.265 ~{GM \over r_c} ~\theta 
\end{equation}
and similarly the scale for potential in the $J_m$ model is $ 0.462 ~G M/r_s$. 
The density scales as $({1/4 \pi a}) ~ M/r_c^3= 0.021 ~M/r_c^3$ for the $r_c$ 
model whereas it scales as $ 0.0368 ~M/r_s^3$ for the $J_m$ model. The radial
orbit period can be written as 
\begin{equation} 
T_r = {2 r_c \over \sqrt{\alpha}}
\int_{\xi_1}^{\xi_2} {\diff \xi \over \sqrt{ -\cE-\Phi(\xi)-J^2/2 \xi^2}}
\end{equation} 
and hence we define a unit, $T_c \equiv 2 r_c/\sqrt{\alpha} = 2
\sqrt{a}~\sqrt{r_c^3/GM}$. For the $r_c$ model, $a= 3.38 ~(T_c=3.68)$ and $a= 
2.16 ~(T_c=2.94)$.

\noindent {\it The density profile}

For large $r$, the densities, $\rho(r)$, for both the models scale as
$r^{-4}$; this has been derived analytically in appendix \ref{e1.5} and is also 
apparent from Fig. \ref{rho}.  The density has
a sharp break at $r=r_c$ for the $r_c$ models. The fraction of mass in the
core, is about 5\% for both models. The density continues to decrease
gently with $r$ upto about $5 r_c$, beyond which the slope changes rather
abruptly to a much steeper value, and rapidly converges to the asymptotic
$r^{-4}$ profile (see the log-log plot in the lower panel of
Fig.~\ref{rho}). The radius at which the break occurs is a few $r_s$ (see
eqn (\ref{rbc})) for both models. This is seen in the C runs of
van Albada (1982), see Fig 6 in the paper. A similar break is also seen in
the cosmological simulations of Tormen, Bouchet \& White (1997) and Moore
et al. (1998) where the power law is roughly $r^{-1}$ for the inner region
and between $r^{-3}$ and $r^{-4}$ for the halo. 
 \placefigure{rho}

\noindent {\it The differential energy distribution}

Here we use $E=-\cE$ in order to compare our results with simulations. The
differential energy distribution, $N(E)$, is a useful indicator of the
phase space structure.  The differences in the analytic forms of $N(E)$
for the different constraints are indicated by eqns (\ref{nepc}, \ref{nerc}, \& 
\ref{nejm}) which corroborate each other in the
Keplerian limit (this is a reflection of the shape of the ${\cal C}(\eta)$
in the $\cE-J^2$ plane). Whereas the real space structure (especially at
large $r$) for the two models are similar, the plots in Fig.~\ref{nefig}
demonstrate how different the models are at high energies; $N(E)$ for the
$J_m$ model rises sharply, reaches a maximum at some $E$, and thereafter
is independent of $E$.  For the $r_c$ model, $N(E)$ is smooth, rising
gradually, and becomes a strictly {\em linear} function of $E$ (see Fig 7
of van Albada 1982, runs C2 and C3 for a linear plot). The linear region
in our model begins at energies 0.7 of the well depth (c.f appendix\ref{e1.5}),
i.e, it is dominantly linear and is consistent with C2 and C3. The log
plot, Fig.~\ref{nefig} is also consistent with Fig 4-20 in Binney \&
Tremaine of run C3 and the simulations of Spergel \& Hernquist (1992),
their Fig. 2. Also it is clear from eqn (\ref{jeta}) that the non-zero
intercept of the constraint, ${\cal C}(\eta)$, which is $J^2_\eta(\cE=0)$,
has the desirable effect of implementing Jaffe's (1987) insight that the
differential energy distribution $N(E)$ tends to a non-zero value as $E
\rightarrow 0$ from below, since the probability of ejection from the core
is not expected to be sensitive to small changes in the final energy.

\placefigure{nefig}

\noindent {\it De Vaucoleurs $r^{1/4}$ law}

In both cases, the surface densities provide reasonably good fits to the de 
Vaucoleurs $r^{1/4}$ law 
in the range 0.1--8 $r_e$ as indicated in Fig.~\ref{sigma}, assuming a constant 
mass-to-light ratio. This includes the range in radii (0.1 --2 $r_e$) over 
which 
the $r^{1/4}$ law provides excellent fits to the brightness profiles of 
ellipticals (Burkert~1993). The slope in the figure is close to $-8$ (the 
standard slope is -7.67) where $\Sigma$ was normalized to its value at the 
center unlike the usual normalization by the extrapolated peak value.
\placefigure{sigma}

\noindent {\it The anisotropy profile}

The DF, eqn (\ref{us}), naturally gives radially anisotropic models where most 
of
the mass is outside the core radius.  The anisotropy profile and velocity 
dispersions of the $\cE^{3/2}$ models which is very similar to that of the $T_r$ 
models.  Fig. \ref{beta} shows a plot of the run of the anisotropy parameter, 
$\beta$, with radius. Both the $r_c$ and $J_m$ models are very nearly isotropic 
within the core and rapidly become anisotropic. The Fig. \ref{vr2} shows the 
radial velocity dispersions which indicate that the $r_c$ model is slightly more 
anisotropic than the $J_m$ model. The dispersion and anisotropy profiles are 
very similar to the Fig 8 of van Albada (1982).
\placefigure{beta}
\placefigure{vr2}

\section{Discussion}
\label{disc}

In \S \ref{quasi} we made an assumption, based on a plausible picture of
diffusion in action space, that $A(E)={\rm constant}$; as discussed in the
previous section, the agreement with simulations is encouraging.
However, the physical
picture of mixing without friction leads to a 
constant $A$ only if the constraints are hard and the mixing lasts enough 
time, two things we cannot lay down a priori. So the success of the
simplest such model does teach us that this picture  is a good first
approximation, possibly improvable.

We now compare and contrast our approach with two earlier papers which can
be regarded as fitting into the same broad framework of constrained
violent relaxation.  MTJ explored DFs with Gaussian and Lorentzian cutoffs
in $J$, multiplied by an exponential function of $E$; thus their models
have one more parameter, a temperature, $1/\beta$ (c.f. Stiavelli \&
Bertin 1987). The infinite temperature limits of their models, in common
with our sharply cutoff $J_m$ model, have an asymptotically flat $N(E)$
near $E=0$. In order to obtain an increasing $N(E)$ with the MTJ models,
one requires a negative $\beta$. As discussed earlier, our $r_c$ model
gives an increasing $N(E)$ without the need for an exponential factor. We
therefore find the $r_c$ model is preferable to the $J_m$ model. We note
that even if the cutoff function, ${\cal C} (\eta)$ is not as abrupt as
given in eqn (\ref{cutoff}), $N(E)$ is expected to behave in a similar
fashion.

SH proposed a kinetic model of the wave-particle interaction process, with
an associated variational principle analogous to Boltzmann's H-theorem,
and made a comparison with their own extensive simulations.  Their
description of energy changes occurring by a sequence of kicks is in fact
close to the picture here.  We used this to motivate a {\em constant}
phase space density modulated by a pericenter or angular momentum cutoff
and Tremaine's incompleteness factor $(T_e/T_r)$. In contrast, SH assume
that the kinetics drives the system to the maximum of their entropy
functional.  However, their Boltzmann-like factor contains a {\em
negative} temperature (as in the MTJ models), but this equilibrium is
meaningful only in systems for which the density of accessible states
decreases rapidly with energy (c.f. \S~73 of Landau \& Lifshitz~1980). We
suggest that the SH kinetic picture, with which we are in broad agreement,
would not actually lead to a negative temperature distribution. The more
appropriate physical picture is one of mixing in phase space.

\section{Conclusions and Caveats} 
\label{conclude} 
We have presented a
semi-analytic model of violent relaxation that includes a new picture of 
diffusion in phase space
with a novel implementation of the pericenter constraint, and the $1/T_r$ 
incompleteness
factor.  Notions of even a partial thermal equilibrium with Boltzmann-like
exponential factors play no role -- a property we regard as a virtue in
describing a collisionless system. The rise in the energy distribution
function which such factors (with negative temperatures!) mimicked in
earlier work now arises naturally from the pericenter cutoff in our
calculation. The resulting properties of density, surface brightness,
energy distribution, anisotropy profiles are in good agreement with
simulations.  The fact that the properties of the models are parameter
free ($M$ and $r_s$ are merely used to normalize), may be a considered a
virtue, as they demonstrate that the constraints explored here seem to
capture the essential details in the case of cold non-rotating collapse.
More realistic systems can be obtained if one deviated from the
simplifying assumptions and comprehensively explored associated models
such as $1/(T_r +T_0)$ in \S \ref{1/tr}, albeit with more parameters. Even
though sophisticated numerical codes that now exist to perform N-body
experiments, semianalytic models help in understanding their output. They
also have the advantage of fitting simulations and observed galaxies
reasonably well. Possible directions of future work include a more
detailed comparison of our models with numerical simulations,
investigations of stability and extension to axisymmetric systems.

\appendix
\section{$\cE^{3/2}$ model with pericenter cutoff}
\label{e1.5}
We have carried out a detailed study of models with $f_0$, the first
factor in eqn (\ref{us}), chosen to be $\cE^{3/2}$, with both kinds of
cutoff, viz. pericenter and angular momentum. Exisiting work, by MTJ, uses
a gaussian rather than sharp cutoff in angular momentum. The pericenter cutoff 
has not been implemented previously and \S \ref{ftr}  gives some details to 
enable the interested reader
to see how  the sharply cutoff models are constructed, and other details of its 
properties along with comparisons to simulations can be
found elsewhere (Mangalam \& Sellwood 1999).
As a first  approximation and the simplest model,  we consider  the distribution 
function $f_0=\cE^{3/2}$, with a sharp pericenter cutoff at $r_c$. Then the 
integral in \S \ref{ftr1} can be written as 
\begin{eqnarray}
{\cal I}_0(r; \cE^{3/2})&= &{1 \over \sqrt{2} r^2} 
\int_0^{J^2_e} \diff J^2 \int_0^{B(r,J^2)} \diff \cE ~\cE^{3/2} (-\cE +B)^{-1/2} 
\nonumber  \\
&=&{\pi \over \sqrt{8}} (-\Phi)^3
\end{eqnarray}
Similarly, the integral, 
\begin{equation}
{\cal I}_-(r;\cE^{3/2}) \equiv {\cal I}_-=  {\pi \over \sqrt{8}} \cE_\ast^3
\end{equation}  
but where  $J^2_\ast$ is the lower limit to $J^2$ and 
\begin{equation}
{\cal K}_-(r;\cE^{3/2})=  {\pi \over \sqrt{8}} (-\Phi-{J_m^2 
\over 2 r^2})^3
\end{equation}
where $J^2_m$ is the lower limit to $J^2$.
 Now the ``$+$" integral works out to be
\begin{eqnarray}
{\cal I}_+(r;\cE^{3/2})&=& {1 \over r^2}\int^{\cal E_\ast}_0 
\diff \cE ~ \cE^{3/2}\left. \sqrt{(-\Phi -\cE) 2 r^2 -J^2}\right ]_{ 
J_\ast^2}^{-(\cE+\Phi_c)r_c^2} \nonumber \\
&=& {\pi \over \sqrt{8}} \cE_\ast^3 (1-\sqrt{(r^2-r_c^2)/r^2})
\end{eqnarray}
Gathering all the integrals in $\cal I$, defined above and (\ref{i}), and  
absorbing the numerical factor into a {\it positive} constant $K$,  Poisson's 
equation can be written as
\begin{equation}
{1 \over r^2}{\diff\over \diff r} (\Phi' r^2)  = K \left \{ \begin{array}{ll} 
-\Phi^3  & r\leq r_c \\ {(r^2-r_c^2)^{-5/2} \over r} (\Phi r^2 -\Phi_c r_c^2)^3  
 -\Phi^3 & r>r_c \end{array} \right.
\label{dense}
\end{equation}
This model is a polytrope of index 3 for $r<r_c$ and we recover the standard 
results  for a distribution function of the form $\cE^{3/2}$ without the 
pericenter cutoff (See Binney \& Tremaine 1987, eqn (4-108c)). For $r>r_c$, the 
contribution to the density is dominated more by  orbits with largely radial  
velocities and the first of the 2 terms limits the orbits to those which have 
angular momenta less than the bound specified by ${\cal A}_2$. In the limit of 
$r_c=0$, the RHS of (\ref{dense}) for $r>r_c$ vanishes as expected. We now scale 
the radius by $\lambda =r/\xi$, and the potential by $\alpha =\Phi/\theta$, 
similar to scalings employed for the polytrope problem  (with the difference 
that we have the normalization constant $K$) and obtain the following relations
\begin{equation}
K \alpha^2 \lambda^2 =1
\end{equation}
\begin{equation}
 GM /(\lambda  \alpha) =a \equiv  -\int_0^{\xi_c} \xi^2 \theta^3  ~\diff  \xi 
-\int_{\xi_c}^\infty \xi \left \{-(\xi^2-\xi_c^2)^{-5/2} (\theta \xi^2 -\theta_c 
\xi_c^2)^3   + \xi \theta^3 \right \} \diff \xi
\label{norm} 
\end{equation}
where $a$ is a geometric factor that depends on the solution. Here we pick 
$\lambda/r_c=\xi_c=1$, (so that $\alpha={GM \over a~ r_c }$ and $K= ({a \over 
GM})^2$ ) such that the cutoff radius in these units is unity and it simplifies
the task of finding the solution. This model has two parameters,  the total mass 
$M$ and the cutoff radius $r_c$ and a unique solution can be found from the 
resulting equation, where  $\theta_c=\theta(1) \equiv \theta_1$
\begin{equation}
{1 \over \xi^2}{\diff\over \diff \xi} (\theta' \xi^2)  = \left \{ 
\begin{array}{ll} -\theta^3 & \xi\leq 1 \\ {(\xi^2-1)^{-5/2} \over \xi} (\theta 
\xi^2 -\theta_1 )^3  -\theta^3 & \xi >1 \end{array} \right.
\label{theta}
\end{equation}
with the boundary conditions
\begin{eqnarray}
\theta'(0)=0 ~~\& && \theta(\infty)=0.
\label{bc}
\end{eqnarray}
The former is a consequence of an implicit assumption of a non-existence of a 
central point mass and the latter boundary condition is enforced to be 
consistent with the distribution $f_0=\cE^{3/2} \sim 1/T_r$ vanishing at 
$\cE=0$. It is interesting to study the asymptotics of this model by changing to 
a convenient variable $u=1/\xi$. For $ u <1$, we obtain
\begin{equation}
u^4 \theta''(u)= (\theta -\theta_1 u^2)^3 (1-u^2)^{-5/2} -\theta^3
\label{u} 
\end{equation}
As $r \rightarrow \infty$ or near $u=0$, we can expand in powers of $u$ to 
obtain
\begin{equation}
u^4 \overline{\theta}''(u)/\theta_1^2=-3 u^2 \overline{\theta}^2 +3 
\overline{\theta} u^4 +(5/2) \overline{\theta}^3 u^2 -u^6+O(u^7)
\end{equation}
where $\overline{\theta}=\theta/\theta_1$, which leads to the following solution 
of $\theta(u)$ near $u=0$
\begin{equation}
\overline{\theta}(u)= c_1 u-(3/2) c_1^2 \theta_1^2 u^2 +O(u^3)
\end{equation}
where $c_1=a/\theta_1$ since $\Phi \rightarrow -GM/r_c~u$ or $\theta  
\rightarrow -a ~u$. 
As a result the density, $\rho$,  in units of $M/r_c^3$ asymptotically behaves 
as
\begin{eqnarray}
\rho = u^4 \theta''(u)&=&\theta_1^3 \left \{-3  c_1^2 u^4 + \left [3 c_1 -9 
c_1^2 \theta_1^2 +(5/2) c_1^3 \right ] u^5 \right \} +O(u^6).
\end{eqnarray}
Hence, as $r \rightarrow \infty$, $\rho \propto r^{-4}$. Now the radius beyond 
which the leading term dominates, which we call as the break radius, is defined 
as
\begin{equation}
r_b \simeq r_c \left |{1 \over c_1} +{5 \over 6} c_1-3 \theta_1^2 \right |
\label{rbc}
\end{equation}

Next, we solve the equations, (\ref{theta}) and (\ref{bc}). It is convenient to
use Lane-Emden solutions in the region, ($0\leq\xi\leq 1$), and use (\ref{u}) in 
the region $1 >u \geq 0$, with the boundary condition $\theta(u=0)=0$. A 
sufficiently accurate solution for the region $0\leq\xi\leq 1$ can be obtained 
by power series expansions. The solution upto order 12 is given in terms of the 
well depth $\theta(0)=\theta_0$ can be obtained from power series expansions.
Using that solution for a trial well depth, $\theta_0$, $\theta(u=1)$ and 
$\theta'(u=1)=-\theta'(\xi=1)$ are calculated and  (\ref{u}) is numerically 
integrated to $\theta(u=0)$. A value of $\theta_0$ is for which the boundary 
condition, $\theta(u=0)=0$ is satisfied is determined iteratively. This occurs 
for 
\begin{equation}
\begin{array}{lcr}
\theta_0=-0.821 & \theta'(u=1)=-0.15264 & \theta(u=1)=-0.737234
\end{array}
\end{equation} 
and the corresponding $\theta(\xi)$ is shown in Fig. \ref{phi}. Note that for 
this solution, $a=3.38$.
\placefigure{phi}

\figcaption[Fig1.eps]{The two possible locations of $r_c$, $r_{c1} < r_0$ and $r_{c2} >r_0$ are indicated in the figure where $r_0$ is the circular orbit radius. The allowed orbits are those for which $ \cE \leq \cE_f(r_c, r_0)$, as indicated in the figure, and  $\cE \leq \cE_f(r_0, r_0)$, the energy of the circular orbit, respectively.
\label{ro}}

\figcaption[Fig2.eps]{The applicable bound for $r<r_c$ is the circular orbit energy, $\cE_f(r_0,r_0)$ and for $r>r_c$, it given by $\cE < \cE_f(r_c,r_0)$ as explained in Fig. 1. Lines of turning points are drawn for $r>r_c$, $r<r_c$ and $r=r_c$.
\label{ej2}} 

\figcaption[Fig3.eps]{The areas of integration enclosed by the bold lines-the triangle shaped ${\cal A}_1$, for $r< r_c$ and the wedge shaped ${\cal A}_2$, for $r>r_c$ are indicated in the figure. The limits used in evaluating the $\cal I$ integrals, $\cE_\ast, J^2_\ast$ and $J_m$ are also indicated. The $\cE$ intercept is $-\Phi(r)$ and the foot of ${\cal A}_2$ is $-2 \Phi_c r_c^2$.
\label{wedge}} 

\figcaption[Fig4.eps]{The solution for the potential in the $\cE^{3/2}-r_c$ model is shown in the  figure in units of 0.265 $G M/r_c$ where $r_c$ is the unit of radius.
\label{phi}}

\figcaption[Fig5.eps]{$\Jm(\cE,r)$ for the case $\cE > \cE_c$ and $r> r_c$ is at the point of intersection.
In this case, $\cE > \cE_\ast$, since $\cE_c >\cE_\ast$. This implies that $\Jm=J^2_e=-2(\cE+\Phi) r^2$. When $r<r_c$, the pericenter constraint is satisfied and hence $\Jm=J^2_e$.
\label{ec}}

\figcaption[Fig6.eps]{$\Jm(\cE>\cE_c,r)$ for $\cE_\ast> \cE$ and $\cE_\ast < \cE$ are shown in the figure. $\cE_c >\cE > \cE_{\ast 2} $ for a radius $r=x_2 >r_c$ and $\Jm=-2(\cE+\Phi) r^2$, given by the point of intersection at the energy $\cE>\cE_c$. Similarly, $\cE < \cE_{\ast 1} < \cE_c$  for $r=x_1 >r_c$  and  $\Jm=-2(\cE+\Phi_c) r_c^2$. Also $r_c< x_1 < r_I <x_2$ where $r_I$, given by $(\cE-\Phi_I) r_I^2= (\cE-\Phi_c) r_c^2$, is the apocenter of an orbit with energy $-\cE$ and pericenter $r_c$.
\label{x1x2}}

\figcaption[Fig7a.eps, Fig7b.eps]{The top panel shows a contour plot of the variation of $T_r \cE^{3/2}$ in the allowed $\cE-J^2$ plane. The flat contours indicate that $T_r$  is very nearly isochronic and close to  $\cE^{-3/2}$. The deviation is depicted in the lower panel which shows a section  taken at $J^2/J^2_{cir}=0.5$, where $J^2_{cir}(\cE)$ is the angular momentum of a circular orbit at a given energy, $\cE$.
\label{tre1.5}}

\figcaption[Fig8a.eps, Fig8b.eps]{The upper panel shows the density  as a function $r/r_s$, where $r_s=r_c$ 
for the $r_c$ models, and $r_s$ is the scale radius of the $J_m$ models. The former shows  a sharp drop at $r_c$ due to the pericenter 
constraint whereas the latter has a smooth  profile. The log-log plot  in the lower 
panel shows a break near 5 $r_c$ and an asymptotic form of $r^{-4}$ from 10 --$100~r_c$ 
for both models. The density is shown in units of $0.021~M/r_c^3$ for the $r_c$ model and $0.0368 ~M/r_s^3$ for the $J_m$ model.
\label{rho}}

\figcaption[Fig9.eps]{The $N(E)$ for the $J_m$ model rises abruptly and then is flat, whereas the $N(E)$ for 
the $r_c$ model increases gradually. The well depth in both cases was taken to be 1.
\label{nefig}}

\figcaption[Fig10.eps]{The surface density, $\Sigma$ is an excellent fit to the  $r^{1/4}$ law for  $0.1 < r/r_e < 8$.
$r_e=6.65 r_c$ for the $r_c$ models (lower curve),  and  $r_e=4.03 r_s$ for the $J_m$ 
models (upper curve).
\label{sigma}}

\figcaption[Fig11.eps]{A plot of the anisotropy parameter, $\beta(r) \equiv 1-\overline{v_t^2} /(2 ~\overline{v_r^2})$, for both models. The core is nearly isotropic and becomes nearly  radially anisotropic at 10 $r_s$. The $r_c$ model (upper curve) is slightly more anisotropic than the $J_m$ model (lower curve).
\label{beta}}

\figcaption[Fig12.eps]{ A plot of the normalized radial velocity dispersion, $\sigma_r^2(r) = \overline{v_r^2}(r)/ \overline{v_r^2}(0)$, for both models. The $r_c$ model (lower curve) is slightly more radially anisotropic than the $J_m$ model (upper curve). \label{vr2}}


\begin{thebibliography}{}

\bibitem[]{}Binney, ~J., \& Tremaine, ~S., 1987. Galactic Dynamics, 
Princeton, Princeton University Press.

\bibitem[]{}Burkert, ~A., 1993, Astr. Astrophys., 278, 23

\bibitem[]{}Dejonghe, ~H., 1987, \apj, 320, 477

\bibitem[]{}Jaffe, ~W. 1987. In: Structure and Dynamics of Elliptical 
Galaxies, p. 511, ed. De Zeeuw, ~T., Reidel, Dordrecht.

\bibitem[]{}Landau, ~L. D. 1937, Zh. Eksper. Theor. Fiz. 7, 203.

\bibitem[]{}Landau, ~L. D., \& Lifshitz, ~E. M. 1980. Statistical Physics, 
Oxford, Pergamon.

\bibitem[]{}Litchenberg, ~A. J. \& Lieberman, ~M. A. 1983. Regular and 
Stochastic Motion, New York, Springer.

\bibitem[]{}Lifshitz, ~E. M., \& Pitaevskii, L. P. 1981. Physical 
Kinetics, Oxford, Pergamon.

\bibitem[]{}Lynden-Bell,~D., 1967, MNRAS, 136, 101

\bibitem[]{}Mangalam, ~A., \& Sellwood, ~J. A. 1999, in preparation.
\bibitem[]{}Merrit, ~D., Tremaine,~S., \& Johnstone, ~D., 1989, MNRAS, 236,
 829 (MTJ)

\bibitem[]{} Moore, ~B., Governato, ~F., Quinn, ~T., Stadel, ~J. \& Lake, ~G., 
1998, \apj, 499, L5

\bibitem[]{}Spergel, ~D. N. \& Hernquist, ~L., 1992, \apj, 397, L75 
(SH)

\bibitem[]{}Sridhar, ~S. 1987, J. Astrophys. Astron., 8, 257

\bibitem[]{}Stiavelli, ~M., \& Bertin, G., 1987, MNRAS, 229, 61

\bibitem[]{}Tormen ~G., Bouchet ~F. R., \& White, ~S. D. M., 1997, MNRAS, 286, 
865

\bibitem[]{}Tremaine,~S., Henon,~M., \& Lynden-Bell,~D., 1986, MNRAS, 219,
 285
\bibitem[]{}Tremaine,~S., 1987. In: Structure and Dynamics of Elliptical 
Galaxies, p. 367, ed. De Zeeuw, ~T., Reidel, Dordrecht.

\bibitem[]{}Van Albada, ~T. S., 1982, MNRAS, 201, 939

\bibitem[]{} Yan'kov, ~V. V., 1994, JETP Lett., Vol 60, No. 3, 171

\end{thebibliography}
\end{document}